\documentclass[pra,superscriptaddress,twocolumn,showpacs,floatfix]{revtex4}
\usepackage{graphics}
\usepackage{graphicx}
\usepackage{epstopdf}
\usepackage{amsmath}
\usepackage{bm}
\usepackage{float}

\begin{document}
\title{Rydberg tomography of an ultra-cold atomic cloud}

\author{M.M. Valado}
\affiliation{INO-CNR, Via G. Moruzzi 1, 56124 Pisa, Italy}
\affiliation{Dipartimento di Fisica {\it E. Fermi},
Universit\`{a} di Pisa, Largo Pontecorvo 3, 56127 Pisa, Italy}

\author{N. Malossi}
\affiliation{INO-CNR, Via G. Moruzzi 1, 56124 Pisa, Italy}

\author{S. Scotto}
\affiliation{Dipartimento di Fisica {\it E. Fermi},
Universit\`{a} di Pisa, Largo Pontecorvo 3, 56127 Pisa, Italy}

\author{D. Ciampini}
\affiliation{INO-CNR, Via G. Moruzzi 1, 56124 Pisa, Italy}
\affiliation{Dipartimento di Fisica {\it E. Fermi},
Universit\`{a} di Pisa,  Largo Pontecorvo 3, 56127 Pisa, Italy}
\affiliation{CNISM UdR Pisa, Dipartimento di Fisica {\it E. Fermi}, Universit\`{a} di Pisa, Largo Pontecorvo 3, 56127 Pisa, Italy}

\author{E. Arimondo}
\affiliation{INO-CNR, Via G. Moruzzi 1, 56124 Pisa, Italy}
\affiliation{Dipartimento di Fisica {\it E. Fermi},
Universit\`{a} di Pisa, Largo Pontecorvo 3, 56127 Pisa, Italy}
\affiliation{CNISM UdR Pisa, Dipartimento di Fisica {\it E. Fermi}, Universit\`{a} di Pisa, Largo Pontecorvo 3, 56127 Pisa, Italy}

\author{O. Morsch}
\affiliation{INO-CNR, Via G. Moruzzi 1, 56124 Pisa, Italy}
\affiliation{Dipartimento di Fisica {\it E. Fermi},
Universit\`{a} di Pisa, Largo Pontecorvo 3, 56127 Pisa, Italy}

\begin{abstract}

One of the most striking features of the strong interactions between Rydberg atoms is the dipole blockade effect, which allows only a single excitation to the Rydberg state within the volume of the blockade sphere. Here we present a method that spatially visualizes this phenomenon in an inhomogeneous gas of ultra-cold rubidium atoms. In our experiment we scan the position of one of the excitation lasers across the cold cloud and determine the number of Rydberg excitations detected as a function of position. Comparing this distribution to the one obtained for the number of ions created by a two-photon ionization process via the intermediate $5P$ level, we demonstrate that the blockade effect modifies the width of the Rydberg excitation profile. Furthermore, we study the dynamics of the Rydberg excitation and find that the timescale for the excitation depends on the atomic density at the beam position.

\end{abstract}


\maketitle

Rydberg excitations in ultra-cold gases have been studied extensively in recent years \cite{Gallagher1,Gallagher2,Jaksch,Lukin,Urban,Gaetan}. In particular, the dipole blockade effect \cite{Comparat}, whereby the excitation of one atom to a Rydberg state prevents other atoms within the blockade radius from being excited, has attracted much attention \cite{Vogt1,Tong,Singer,Schauss,Dundin}. One of the main motivations for studying this effect is the application of the dipole blockade in quantum information processing \cite{Saffman,Brion,Muller}. The excitation of Rydberg atoms in the dipole blockade regime was realized in disordered clouds of cold atoms \cite{Tong,Singer,Afrousheh,Cubel, Vogt2,Heidemann1,Diz,Reinhard}, in a Bose-Einstein condensate \cite{Heidemann2,Viteau1} and also using two individual atoms \cite{Urban,Gaetan}. Recently the creation of entanglement in the strongly blockaded regime was demonstrated \cite{Wilk,Isenhower}.

\begin{figure}[H]
\centering
\includegraphics[width=0.44 \textwidth]{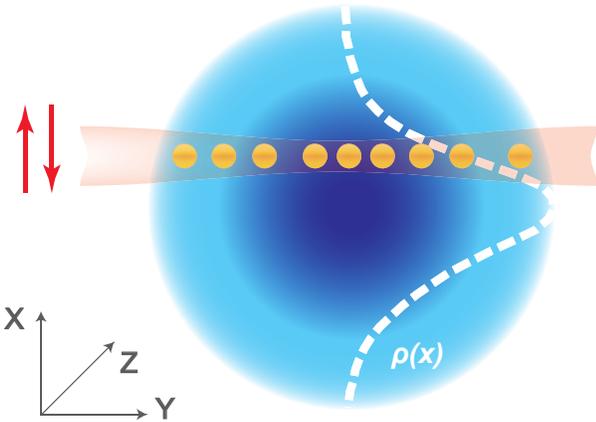}
\caption[1]{(Color online) Experimental procedure for the Rydberg tomography of a cold atomic cloud. The vertical position of the tightly focussed blue excitation laser is scanned vertically across the density profile of the atomic cloud. The larger excitation beam at $1013\,\mathrm{nm}$ is not shown here.}
\label{fig1}
\end{figure}
Here we report the results of an experiment aimed at visualizing the dipole blockade effect in a spatially inhomogeneous cold cloud. In our experiments we create small clouds of ultra-cold 87-Rb atoms in a magneto-optical trap (MOT). The parameters of the MOT (loading flux from a primary two-dimensional MOT, magnetic field gradient, and size of the MOT beams) are chosen such as to obtain clouds with a Gaussian integrated density profile:
\begin{center}
\begin{equation}
\displaystyle \rho(x)=\displaystyle \frac {N} {\sigma_x \sqrt{\pi/2}}e^{-\frac{2x^2}{\sigma_{x}^2}}
\end{equation}
\end{center}
of width $\sigma_x=\sigma_y =\sigma_z\approx 30\,\mathrm{\mu m}$ containing up to $N\approx10^5$ atoms. The resulting peak density, $n_{0}=\displaystyle \frac {2\rho(0)} {\pi\sigma_y\sigma_z}$$\approx 2.5\times 10^{10}\,\mathrm{cm}^{-3}$, corresponds to a mean inter-particle spacing of around $3.5\,\mathrm{\mu m}$. Two excitation lasers at $420\,\mathrm{nm}$ and $1013\,\mathrm{nm}$, with powers $0.75\,\mathrm{mW}$ and $110\,\mathrm{mW}$ and Gaussian waists $6\,\mathrm{\mu m}$ and $80\,\mathrm{\mu m}$, respectively, are then used to excite the atoms to Rydberg states in a coherent two-step excitation scheme (the first step laser at $420\,\mathrm{nm}$ is detuned by $2\,\mathrm{GHz}$ from the intermediate $6P$ level which, therefore, has negligible probability of being populated) with effective two-photon Rabi frequencies up to $400\,\mathrm{kHz}$. After the excitation pulse of duration between $0.2$ and $60\,\mathrm{\mu s}$ an electric field is applied for $2\,\mathrm{\mu s}$ in order to field ionize the Rydberg atoms and to accelerate the resulting ions towards a channeltron. The overall detection efficiency is $\eta\approx 40\%$. During the entire excitation and detection sequence the MOT beams are switched off.
Using a slightly modified version of the above procedure, we can also directly photoionize atoms by leaving the MOT trapping beams switched on and using a $1\,\mathrm{\mu s}$ pulse of the laser at $420\,\mathrm{nm}$ to ionize atoms from the $5P$-level used for laser cooling. The ions thus produced are then detected using the same electric field pulses as in the Rydberg detection.

\begin{figure}[H]
\centering
\includegraphics[width=0.44 \textwidth]{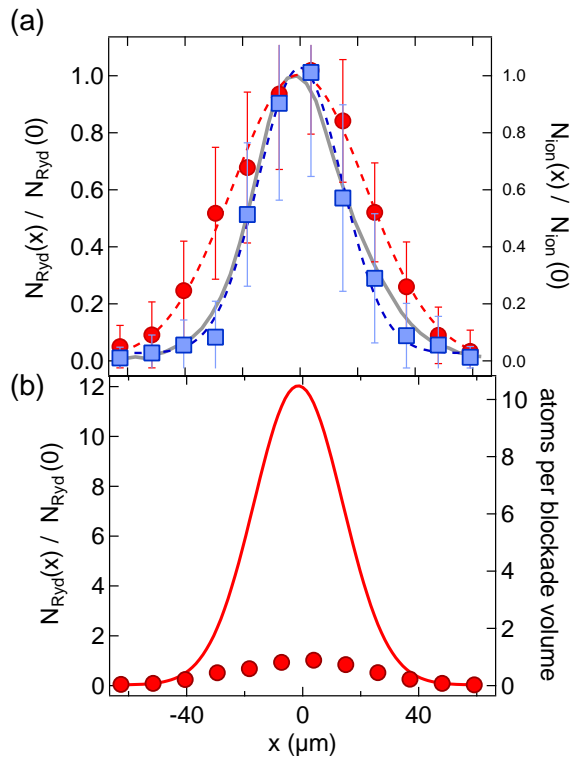}
\caption[1]{(Color online) Spatial tomography of a MOT using Rydberg excitations to the $71D_{5/2}$  state. In (a), the normalized number of detected ions is plotted for the direct ionization from the $5P$ state using only the $420\,\mathrm{nm}$ beam (blue squares) and for the Rydberg excitation (red circles). The duration of the excitation pulse is $1\,\mathrm{\mu s}$ for direct ionization process, in which $N_{Ions}(0)\approx 8$, and for the Rydberg excitation it is $3\,\mathrm{\mu s}$ with $N_{Ryd}(0)\approx 12$. The error bars correspond to one standard deviation of each set of data. For comparison, the horizontally integrated density profile of the MOT is also shown (grey line). Dashed lines are Gaussian fits to the data. In (b), the normalized ion number for the Rydberg excitation plotted in (a) (without the error bars) is shown together with a Gaussian fit to the wings of that distribution with a fixed width equal to that of the MOT (solid line, left-hand side scale). The solid line also represents the spatial variation of the calculated number of atoms per blockade volume (right-hand side scale).}
\label{fig2}
\end{figure}

For both procedures, in order to obtain a good estimate of the mean number of ions and the variance, for each set of parameters the experimental sequence is repeated $50$ times.
To reveal the effect of the dipole blockade on the Rydberg excitation, we probe regions of different densities by scanning the position of the tightly focused blue laser beam at $420\,\mathrm{nm}$ vertically through the density profile of the MOT (see Fig.(\ref{fig1})) using a mirror equipped with a piezo-driven motorized actuator that allows us to control the position of the laser beam in steps of $0.7\,\mathrm{\mu m}$ (a similar technique was demonstrated recently in \cite{Lochead}).

Since the MOT has a roughly Gaussian shape in all three spatial directions, and the size of the $420\,\mathrm{nm}$ laser beam is much smaller than the MOT size, we can assume that the atomic density is roughly constant over the radial extent of that beam and has a Gaussian shape along its direction of propagation. Therefore, a variation of the position of the laser beam results in a variation of the mean atomic density inside the excitation volume defined by the beam width, while the effective length of the sample along the propagation direction of the beam remains constant. In the case of ionization from the $5P$ excited state (with the MOT trapping beams switched on), the number of ions detected by the channeltron is expected to be directly proportional to the density profile of the MOT in the vertical direction. This is confirmed by the comparison between the ion signal as a function of position and the integrated density profile of the MOT obtained with a CCD camera, in Fig.\ref{fig2} (a).
Repeating the scanning experiment with a two-step Rydberg excitation, we again obtain a roughly Gaussian dependence of the mean detected ion number (and hence the mean number of excitations) on the beam position, but this time the measured width of the distribution is noticeably larger. This can be understood in terms of the dipole blockade: In the high-density region close to the centre of the atomic density distribution, the interactions between the Rydberg states lead to a suppression of the number of excited atoms compared to the non-interacting case, resulting in a flattening - and, therefore, broadening - of the distribution.
\\Another way of visualizing the blockade effect is to fit a Gaussian curve to the wings of the spatial distribution, where one can assume that interaction effects are negligible and hence the measured Rydberg number is roughly linearly proportional to the local density. Fixing the width of the Gaussian to that of the atomic density distribution but leaving the amplitude as a fitting parameter, one obtains the solid curve in Fig.\ref{fig2} (b). That curve shows the number of Rydberg excitations one expects to find in the absence of the dipole blockade effect based on the numbers observed in the low-density wings of the atomic cloud. At the centre of the distribution, the detected number of Rydberg atoms is suppressed by a factor of $\approx 12$ compared to the expected number.
Fig.\ref{fig2}(b) also shows the number of atoms per blockade sphere as a function of position, calculated assuming a blockade radius of around $10\,\mathrm{\mu m}$ expected for the $71D_{5/2}$ state \cite{Viteau1} used in this experiment. While in the wings of the distribution ($50-60 \,\mathrm{\mu m}$ from the center) that number is less than $\approx 0.1$, justifying the above assumption of negligible interactions, towards the centre of the MOT it rises to around $10$. This number is in good agreement with the suppression by a factor $12$ of the Rydberg excitations calculated above.

\begin{figure}[H]
\centering
\includegraphics[width=0.44 \textwidth]{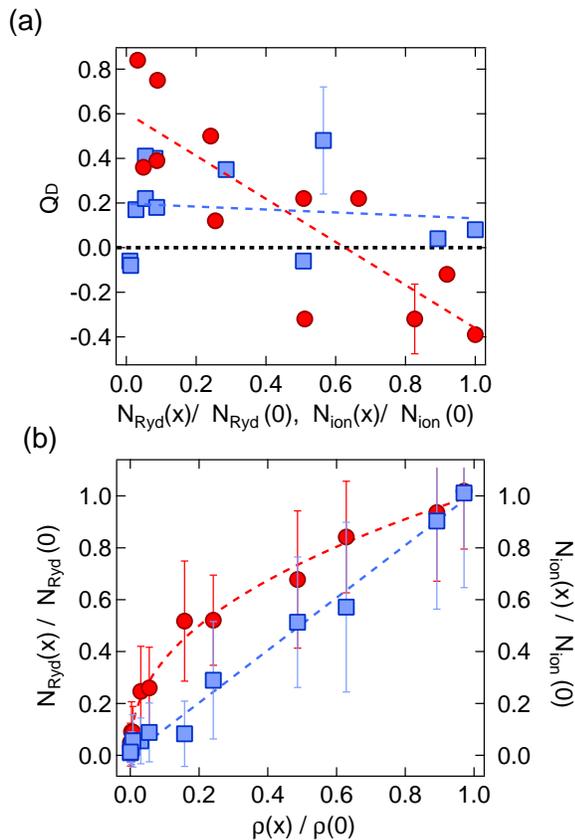}
\caption[1]{(Color online) Characterization of the Rydberg excitations to the $71D_{5/2}$ state as a function of the MOT local density and statistics of the detected ions. In (a), the data of Fig.\ref{fig2}(a) is plotted as a function of the normalized local density in the MOT. The Rydberg excitation (red circles) data are fitted with a power law with exponent $0.43\pm 0.02$ whereas a linear function is used to fit the data corresponding to direct ionization (blue squares). In (b), the Mandel $Q$-factor of the detected ion signal for both experiments is shown as a function of the normalized number of detected ions. Typical error bars (one standard deviation) for the blue and red symbols are shown. Dashed lines are linear fits to guide the eye.}
\label{fig3}
\end{figure}

The suppression of excitations due to the dipole blockade is also clearly visible in a plot of the ion and Rydberg numbers as a function of the local density in the MOT, as shown in Fig.\ref{fig3} (a). Whereas the number of ions depends linearly on the density, the dependence of the number of Rydberg atoms on the density exhibits a decreasing slope as the density increases.

Further evidence for the blockade effect is obtained by measuring the fluctuations in the number of detected ions. In the regime in which the interactions between atoms are negligible, each excitation to a Rydberg state is independent of all the others, leading to a Poissonian excitation process.

In the strongly interaction regime (i.e., for high density and / or large blockade radius), however, the excitation dynamics of the atoms are strongly correlated \cite{Viteau2} and one expects sub-Poissonian counting statistics. Both processes can be quantified through the Mandel-$Q$ parameter given by:

\begin{figure}[H]
\centering
\includegraphics[width=0.38 \textwidth]{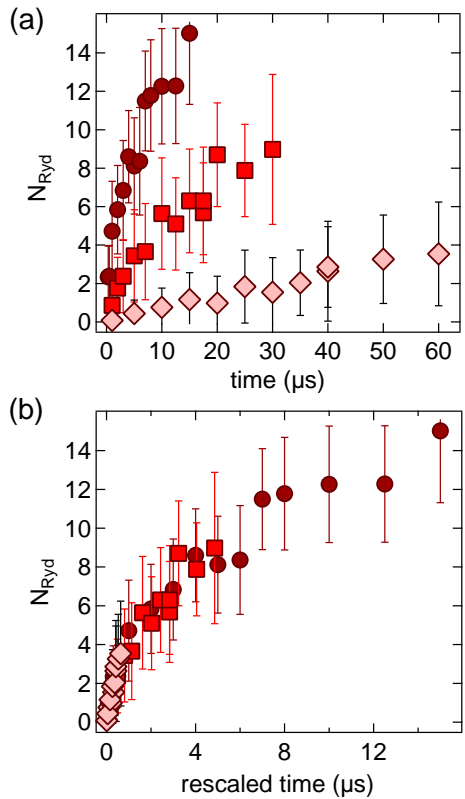}
\caption[1]{(Color online) Excitation dynamics in the $88D_{5/2}$ Rydberg state for different beam positions: at the center of the MOT (dark red circles) and at distances of $1\sigma_x$ and $1.5\sigma_x$ from the MOT center (medium red squares and light red diamonds, respectively). Fig.\ref{fig4} (a) shows the mean number of detected Rydberg atoms as a function of time. In Fig.\ref{fig4} (b) the same data is plotted as a function of a renormalized time that takes into account the different collective Rabi frequencies due to the different local densities. The error bars correspond to one standard deviation of each set of data.}
\label{fig4}
\end{figure}

\begin{center} \begin{equation}
\displaystyle Q= \displaystyle \frac {\Delta N} {\langle N\rangle}-1,
\end{equation}
\end{center}

where $\langle N\rangle$ is the mean number of Rydberg excitations and $\Delta N$ is the variance. For Poissonian processes $Q=0$, while sub-Poissonian processes lead to negative values of the Mandel-$Q$ parameter. The value $Q=-1$ corresponds to a theoretical case in which there is a complete suppression of fluctuations.

The effect of the dipole blockade on the $Q$-parameter is shown in Fig.\ref{fig3} (b), where the detected $Q$-factor, $Q_D=\eta Q$, is plotted as a function of the detected ion signal. For the case of Rydberg excitations (red circles) $Q_D$ depends linearly on the mean number of excitations. The negative values of $Q_D$, reached at high values of the number of detected ions, are a clear indication of sub-Poissonian statistics due to the dipole blockade which leads to strong (anti-) correlation of the excitation probabilities.  In the direct ionization experiment (blue squares), $Q_D$ is approximately independent of the mean number and fluctuates between $0$ and $0.4$ which, given the limited sample size of $50$, is compatible with a Poissonian process for which in practice one expects $Q_D\ge0$ (since technical noise will result in a positive value of $Q_D$).

Finally, in Fig.\ref{fig4} (a) we show the time evolution of the detected number of Rydberg atoms at three different positions of the $420\,\mathrm{nm}$ beam (and, therefore, for three different local densities $\rho(x)$): at the center of the cloud, at $1\sigma_x$ and at $1.5\sigma_x$ from the center, respectively. In the strongly blockaded regime the Rydberg excitations are expected to occur at a collective Rabi frequency proportional to $\sqrt{N}$ , where $N$ is the number of atoms inside a blockade volume \cite{Pfau}. This picture is confirmed by the results shown in Fig.\ref{fig4}. In \ref{fig4}(a) the different timescales for the Rydberg excitation in different density regions of the MOT are clearly visible. When rescaling the excitation time by a factor $\sqrt{\rho(x)/\rho(0)}$, as shown in Fig.\ref{fig4} (b), the three curves of Fig.\ref{fig4} (a) fall on top of each other.

In summary, we have demonstrated a method for spatially visualizing the effect of the dipole blockade in the excitation of strongly interacting Rydberg atoms inside a cold atomic cloud. We have also shown that the Mandel-$Q$ parameter is negative in the high density regions as expected for a sub-Poissonian excitation process due to the dipole blockade.

\
We acknowledge financial support from the EU Marie Curie ITN COHERENCE Network.

\bibliographystyle{apsrmp}

\end{document}